\documentstyle[aps,twocolumn]{revtex}
\begin{document}
\draft
\title{Structure of the Quark Propagator at High Temperature}
\author{H. Arthur Weldon}
\address{Department of Physics,
West Virginia University, Morgantown, West Virginia, 26506}
\date{July 14, 1999}
\maketitle
\begin{abstract}
In the high temperature, chirally invariant phase of QCD, the quark
propagator is shown to have two sets of poles with different dispersion
relations. A reflection property in momentum space  relates all
derivatives at zero-momentum of the particle and hole energies,
the particle and hole damping rates, and the particle and hole residues.
No use is made of perturbation theory.
\end{abstract}


\subsubsection{Background}

In the high temperature, chirally invariant phase of gauge theories the
fermion propagator has some unusual properties in the one-loop approximation.
Despite explicit chiral invariance the fermion has an effective mass
proportional
to temperature
\cite{a1,a2}.  The ``mass" is a consequence of the preferred reference frame of
the heat bath, which allows the pole of the fermion propagator to be off the
light cone. The second and more surprising effect is that there are two
poles in
the propagator corresponding to two different dispersion relations, both with
positive energy \cite{a1,b1,b2}.  At zero momentum the two branches coincide.
The branch describing normal particle
excitations increases monotonically with momentum.  The other branch is a
collective excitation, referred to either as a hole  or as a plasmino,
that decreases slightly at small momentum, reaches an absolute minimum, and
then rises.
The  excitations on the upper branch of the dispersion curve have the same
chirality and helicity (i.e. both positive or both negative) as is
customary for particles. The hole excitations of the lower branch have
chirality opposite to helicity. These results are reviewed by Le Bellac
\cite{b3}.

In the hard thermal loop
approximation of Braaten and Pisarski, the effective fermion propagator is the
starting point for consistent higher order calculations
\cite{c1,c2,c3,c4,c5,c6}.  The doubled fermion dispersion relations have been
  important in calculations of dilepton production \cite{d1,d2,d3} and
strangeness production \cite{e1} by a quark-gluon plasma.
There have been investigations of how the dispersion relations are
affected by retaining
non-leading powers of temperature \cite{f1,f2},
by including a bare mass
\cite{g1,g2}, and by a including a chemical potential
\cite{h1,h2,h3}.

 In the electroweak sector at high  temperature the quarks, charged
leptons, and neutrinos all have  effective thermal masses and  doubled
dispersion relations
\cite{a2,i1,i2,i3,i4,i5}. The phenomena also occurs with Majorana
fermions \cite{i6}.  Farrar and Shaposhnikov exploited the importance of
chirality transport in baryogenesis at the electroweak scale and
used the doubled fermion dispersion relations to account for the baryon
asymmetry of the universe \cite{j1}. However, it has been shown that
including fermion damping  makes the effect too small
\cite{j2,j3,j4,j5}.  There are renewed attempts to account for  baryogenesis at
the electroweak scale using the fermion dispersion relations in models with two
Higgs doublets \cite{j6} and in minimal supersymmetric models \cite{j7}.

\subsubsection{QCD Motivation}

The evidence discussed thus far for  the thermal fermion mass and for the
doubled  dispersion relations is based entirely on one-loop perturbation
theory. The one-loop calculations are sufficiently accurate for electroweak
effects but not for QCD phenomenology. Existing
calculations are only valid if
$g$ is very small. Unfortunately, the most interesting aspects of the
quark dispersion relations are limited both by the small coupling  and also by
some accidentally small coefficients.
For example, the thermal mass is
  $m= 0.41 gT$. The minimum in the dispersion relation for the hole excitation
is  slightly below this at  $E=0.38 gT$
and the minimum occurs at a very small momentum, $p=0.17 gT$.
At temperatures well above the critical temperature $g$ is indeed small, but
near $T_{c}$ it is not.

 There is some nonperturbative evidence that the one-loop calculations are
qualitatively correct. Peshier et al \cite{k1,k2}  and also L\'{e}vai and
Heinz
\cite{k3} have successfully fit
lattice simulations of high temperature QCD  using
 effective quark masses \cite{k1,k2,k3}. These
studies do
 not test the doubled dispersion relation. Recently Sch\"{a}fer and
Thoma have computed the  quark self-energy in the presence of a
gluon condensate
\cite{l1}. Their calculation is one-loop but with the gluon condensate
determining the gluon propagator. In their calculation it is the  soft
loop momenta that control the quark dispersion relation rather than
the hard momenta as is usually the case.
Nevertheless the results are qualitatively similar in several
 respects. There is an effective mass, but now related to the
condensates
$\langle\vec{E}^{2}\rangle$ and $\langle\vec{B}^{2}\rangle$. Using
lattice data for the condensates gives $m\approx 1.15 T$ for temperatures
in the
range $1.1T_{c}\le T\le 4T_{c}$.  There are, in addition, two different
dispersion
relations. They coincide at
$p=0$  but at small $p$ have opposite slopes. The dispersion relation
for the quark is monotonically increasing; that for the hole decreases
slightly, reaching a minimum at $p\approx m/2$  and then
increasing at larger $p$. The self-energy computed
by   Sch\"{a}fer and Thoma
is a polynomial in  $p_{0}$ and $p$ divided by $(p_{0}^{2}-p^{2})^{3}$,
whereas the hard thermal loop self-energy contains $\ln(p_{0}\pm p)$.

This naturally leads to the question of whether the existence of the
separate particle and hole dispersion relation are independent of weak coupling
and of  the one-loop approximation.
This paper will demonstrate that, independently of
perturbation theory, the fermion propagator will always have two
distinct dispersion relations in the high temperature,  chirally
symmetric phase.

\subsubsection{Retarded propagator}

It will be most convenient to analyze the retarded propagator and
later express the time-ordered propagator in terms of it.
The free
retarded propagator is
\begin{displaymath}
S^{R}_{\rm free}(p_{0},\vec{p})={\gamma_{0}p_{0}-\vec{\gamma}\cdot\vec{p}
\over (p_{0}+i\eta)^{2}-\vec{p}^{2}},
\end{displaymath}
and is independent of temperature.
It  can be rewritten as
\begin{equation}S^{R}_{\rm
free}(p_{0},\vec{p})={1\over 2}{\gamma_{0}-\vec{\gamma}\cdot\hat{p}
\over p_{0}-p+i\eta}
+{1\over 2}{\gamma_{0}+\vec{\gamma}\cdot\hat{p}
\over p_{0}+p+i\eta}.\label{Sfree}\end{equation}
The particle pole is in the fourth quadrant of the complex $p_{0}$ plane
at $p_{0}=p-i\eta$; the antiparticle pole is  in the third quadrant at
$p_{0}=-p-i\eta$.

The full retarded propagator has both poles and cuts but the
singularities are always in the lower half-plane.  At nonzero
temperature the propagator in the rest frame of the heat bath will depend
separately on energy $p_{0}$ and momentum $\vec{p}$.  Invariance under
chirality and parity limits the self-energy to be a linear combination of
the matrices $\gamma_{0}$ and
$\vec{\gamma}\cdot\vec{p}$. The most general possibility can be written
\begin{equation}
S^{R}(p_{0},\vec{p})\!=\!{{1\over 2}(\gamma_{0}-\vec{\gamma}\cdot\hat{p})\over
 D_{+}(p_{0},p)}
+{{1\over 2}(\gamma_{0}+\vec{\gamma}\cdot\hat{p})\over
D_{-}(p_{0},p)}.\label{Rpropagator}\end{equation}
The arguments below will show that $S^{R}$ contains four poles.  In the fourth
quadrant there is a pole at a complex energy ${\cal E}_{p}$ for particle
excitations and another at  a different complex energy ${\cal E}_{h}$ for hole
excitations.  The two poles for the corresponding antiparticle and antihole are
in the third quadrant.

\subsubsection{Existence of separate dispersion relations}

 The first step
in the argument is the observation that at $p=0$ the propagator cannot
depend on the
direction of the unit vector $\hat{p}$. This is  because the
dependence on $\hat{p}$ of the fermion self-energy comes entirely from the
the vector $\vec{p}$ as shown in Appendix A.
Therefore  $D_{+}(p_{0},0)=D_{-}(p_{0},0)$ and the propagator has the form
\begin{equation}
S^{R}(p_{0},0)={\gamma_{0}\over 2D_{+}(p_{0},0)}={\gamma_{0}\over
2D_{-}(p_{0},0)}.\label{p=01}\end{equation}
 Now comes the one dynamical input, namely that when $p=0$ there is a pole
in the energy  variable at the thermal mass of the quark so that
\begin{equation}
S^{R}(p_{0},0)={\gamma_{0}\over 2(p_{0}-{\cal
M})\,a(p_{0})},\end{equation}
 where $a(p_{0})$ is analytic
at $p_{0}\!=\!{\cal M}$.  The  mass ${\cal M}$  is complex with a  negative
imaginary part since it is a pole of the retarded propagator.
 Denote this mass by
\begin{equation}
{\cal M}=m-i\gamma/2.\label{mass}
\end{equation}
From dimensional analysis both $m$ and $\gamma$ must be proportional to
temperature. The real part, $m$, has been extracted from lattice calculations
\cite{l2}.

The next step is to examine $p\!\neq\! 0$. The denominators $D_{\pm}(p_{0},p)$
must have the structure
\begin{eqnarray}
D_{+}(p_{0},p)=&&(p_{0}-{\cal M})a(p_{0})-b(p_{0},p)\nonumber\\
D_{-}(p_{0},p)=&&(p_{0}-{\cal M})a(p_{0})-c(p_{0},p)\nonumber,\end{eqnarray}
where  $b$ and $c$ both  vanish at  $p\!=\!0$.
Since $b$ and $c$ are continuous in $p$ and  vanish at zero momentum, they
can be
made arbitrarily small by  choosing $p$ sufficiently small. Therefore in the
neighborhood of
$p_{0}\approx {\cal M}$ and $p\approx 0$ one can approximate
\begin{eqnarray}
D_{+}(p_{0},p)\approx &&(p_{0}-{\cal M})a({\cal M})-b({\cal M},p)\nonumber\\
D_{-}(p_{0},p)\approx&&(p_{0}-{\cal M})a{\cal M})-c({\cal
M},p)\nonumber.\end{eqnarray}
$D_{+}$ vanishes at a complex energy $p_{0}={\cal E}_{p}$ given  by
\begin{mathletters}\begin{equation}
{\cal E}_{p}={\cal M}+{b({\cal M},p)\over a({\cal
M})} +\dots (p\;{\rm small}).\label{dispersion1}\end{equation}
The subscript on ${\cal E}_{p}$ indicates that it is the
complex energy  of the particle excitation. It is, of course, a
function of the momentum $p$.
$D_{-}$ vanishes at
 a different  energy $p_{0}={\cal E}_{h}$ given by
\begin{equation}
{\cal E}_{h}={\cal M}+{c({\cal M},p)\over a({\cal
M})} +\dots (p\;{\rm small}),\label{dispersion2}\end{equation}
\end{mathletters}
 which is
the complex energy of the collective hole excitation. Note that
it is not necessary to assume that $b$ and $c$ are differentiable at $p\!=\!0$.
 Equations
(\ref{dispersion1}) and (\ref{dispersion2}) must have different momentum
dependence because  $b({\cal M},p) \neq c({\cal M},p)$ or else the full
propagator in Eq. (\ref{Rpropagator})  would be completely independent of
$\vec{\gamma}\cdot\hat{p}$.

For general momentum $p$ the particle and hole energies are solutions to
\begin{mathletters}\begin{eqnarray}
0=&&({\cal E}_{p}-{\cal M})a({\cal E}_{p})-b({\cal E}_{p},p)\label{zero1}\\
 0=&&({\cal E}_{h}-{\cal M})a({\cal E}_{h})-c({\cal
E}_{h},p).\label{zero2}\end{eqnarray}
\end{mathletters}
Since these energies are poles of the retarded propagator, they have negative
imaginary parts:
\begin{mathletters}
\begin{eqnarray}{\cal E}_{p}=&&E_{p}(p)-i\gamma_{p}(p)/2\\
{\cal E}_{h}=&&E_{h}(p)-i\gamma_{h}(p)/2.\end{eqnarray}\end{mathletters}
At zero momentum $E_{p}(0)=E_{h}(0)$ and $\gamma_{p}(0)=\gamma_{h}(0)$
as indicated in Eq. (\ref{mass}).
Although the arguments for this depend only on rotational invariance,
perturbative calculations do show these properties.
 Equality of the zero-momentum energies is displayed in the
one-loop calculations \cite{a1,b1,b2,b3,c1,l1}; equality of the zero-momentum
damping rates was found by Braaten and Pisarski \cite{c2} using HTL
resummation.

\subsubsection{Reflection symmetry}

The momentum variable $p$ originally has a positive, real value $|\vec{p}|$.
However, one can extend the functions $D_{\pm}(p_{0},p)$ to more general values
of $p$.  In particular, it is useful to allow $p$ to be real and negative.
Appendix B shows that
\begin{equation}
D_{-}(p_{0},p)=D_{+}(p_{0},-p),\label{reflectD}\end{equation}
or equivalently $b(p_{0},-p)=c(p_{0},p)$. Under the change $p$ to $-p$
in  Eq. (\ref{zero1}), let the solution be $\Omega\equiv{\cal
E}_{p}(-p)$. Then   Eq. (\ref{zero1}) becomes
\begin{displaymath}
0=(\Omega-{\cal M})a(\Omega)-c(\Omega,p).
\end{displaymath}
Comparison with Eq. (\ref{zero2}) shows that $\Omega={\cal E}_{h}(p)$.
This gives the
reflection relation
\begin{equation}
{\cal E}_{p}(-p)={\cal E}_{h}(p).\end{equation}
The real and imaginary parts of this are
\begin{mathletters}\label{reflection}\begin{eqnarray}
E_{p}(-p)=&&E_{h}(p)\\
\gamma_{p}(-p)=&&\gamma_{h}(p).\end{eqnarray}\end{mathletters}
At $p=0$ this coincides with the previous results.
Differentiating with respect to $p$ and then setting $p=0$ gives a new result:
\begin{mathletters}
\begin{eqnarray}
{\partial E_{p}\over\partial p}\Big|_{p=0}
=&&\,-{\partial E_{h}\over\partial p}\Big|_{p=0}\label{slope}\\
{\partial \gamma_{p}\over\partial p}\Big|_{p=0}
=&&\,-{\partial \gamma_{h}\over\partial p}\Big|_{p=0}.
\label{derivative}\end{eqnarray}\end{mathletters}
(There is a caveat that goes with these derivative relations: It is
possible that
the derivatives do not exist. For example,  one cannot rule out the
existence of
  a branch point at $p=0$, though
 there is no reason to expect it.) Peshier and Thoma have obtained Eq.
(\ref{slope}) nonperturbatively using a Ward identity argument \cite{l1a}. In
perturbative calculations the opposite slopes of the real energies is displayed
in the one-loop calculations \cite{a1,b1,b2,b3,c1,l1} and the opposite
slopes of
the damping rates is a property of   the calculations of Pisarski \cite{c3}
and of
Boyanovsky et al \cite{j5} if one extends their answers to $p=0$. Strictly
speaking, the hard thermal loop calculations of the damping rates do not allow
zero momentum but are limited to $p>g^{2}T$.
Repeated differentiation of Eq. (\ref{reflection}) gives the general relations
\begin{mathletters}
\begin{eqnarray}
{\partial^{n} E_{p}\over\partial p^{n}}\Big|_{p=0}
=&&\,(-1)^{n}\,{\partial^{n} E_{h}\over\partial p^{n}}\Big|_{p=0}\\
{\partial^{n} \gamma_{p}\over\partial p^{n}}\Big|_{p=0}
=&&\,(-1)^{n}\,{\partial^{n} \gamma_{h}\over\partial p^{n}}\Big|_{p=0},
\end{eqnarray}\end{mathletters}
provided  the derivatives exist.

The vanishing of   $D_{+}$ at $p_{0}\!=\!{\cal E}_{p}$ and of
$D_{-}$ at $p_{0}\!=\!{\cal E}_{h}$ along with the
reflection symmetry (\ref{reflectD}) means that   the retarded propagator
has the
form
\begin{eqnarray}
S^{R}(p_{0},\vec{p})
={1\over 2}(\gamma_{0}&&-\vec{\gamma}\cdot\hat{p})
\Big({Z_{p}(p)\over p_{0}-{\cal E}_{p}}
-d(p_{0},p)\Big)\nonumber\\
+{1\over 2}(\gamma_{0}&&+\vec{\gamma}\cdot\hat{p})
\Big({Z_{h}(p)\over p_{0}-{\cal E}_{h}}
-d(p_{0},-p) \Big).\label{2pole}\end{eqnarray}
Here $d$ is an unknown function
except for the requirement that it has no singularities in the upper half of
the complex $p_{0}$ plane.
The complex residues satisfy the reflection property
\begin{equation}
Z_{p}(-p)=Z_{h}(p),\end{equation}
which implies
\begin{mathletters}\label{reflectZ}\begin{eqnarray}
Z_{p}(0)=&&Z_{h}(0)\\
{\partial^{n} Z_{p}\over\partial p^{n}}\Big|_{p=0}
=&&(-1)^{n}\,{\partial^{n} Z_{h}\over\partial p^{n}}\Big|_{p=0}.
\end{eqnarray}\end{mathletters}
Equality of the residues and of the first derivatives has been found in
one-loop perturbation theory \cite{a1,b1,b2,b3,c1}.

\subsubsection{Charge conjugation invariance}

The propagator in Eq. (\ref{2pole}) is invariant under chirality and parity by
construction. Appendix C shows that invariance under time reversal requires
 Eq. (\ref{T}), which is automatically satisfied by the above.
However invariance under charge conjugation, Eq. (\ref{C}),  is
not automatic but requires the denominators to satisfy
\begin{equation}
D_{-}(p_{0},p)=-[D_{+}(-p_{0}^{\ast},p)]^{\ast}.\label{CC}
\end{equation}
Imposing this on Eq. (\ref{2pole}) gives
\begin{eqnarray}
S^{R}(p_{0},\vec{p})&&\nonumber\\
={1\over 2}(\gamma_{0}&&-\vec{\gamma}\cdot\hat{p})
\Big({Z_{p}(p)\over p_{0}-{\cal E}_{p}}
+{Z_{h}^{\ast}(p)\over p_{0}+{\cal E}_{h}^{\ast}}-f(p_{0},p)\Big)\nonumber\\
+{1\over 2}(\gamma_{0}&&+\vec{\gamma}\cdot\hat{p})
\Big({Z_{h}(p)\over p_{0}-{\cal E}_{h}}
\!+\!{Z_{p}^{\ast}(p)\over p_{0}+{\cal
E}_{p}^{\ast}}\!+\!f^{\ast}(-p_{0}^{\ast},p)
\Big).\label{R}\end{eqnarray}
The new   $p_{0}$ poles at $-{\cal E}_{h}^{*}$ and at $-{\cal E}_{p}^{\ast}$
are due to the  antiparticles of the hole excitation and particle
excitation, respectively. Because ${\cal E}_{p}$ and ${\cal E}_{h}$ are in the
fourth quadrant, then $-{\cal E}_{p}^{\ast}$ and $-{\cal E}_{h}^{\ast}$
are in the third quadrant. The
function
$f(p_{0},p)$ can have no singularities in the upper-half of the complex
plane. It
must satisfy the reflection property
\begin{equation}
f(p_{0},p)=-f^{\ast}(-p_{0}^{\ast},-p),\label{p=03}\end{equation}
which is observed in the one-loop calculations \cite{a1,b1,b2,b3,c1}.

The matrix structure of the propagator comes from the usual Dirac spinors:
\begin{mathletters}\begin{eqnarray}
{1\over
2}\big(\gamma_{0}-\vec{\gamma}\cdot\hat{p}\big)_{\alpha\beta}
=&&\sum_{s}u_{\alpha}(\vec{p},s)\overline{u}_{\beta}(\vec{p},s)\label{pos}\\
{1\over 2}\big(\gamma_{0}+\vec{\gamma}\cdot\hat{p}\big)_{\alpha\beta}
=&&\sum_{s}v_{\alpha}(-\vec{p},s)\overline{v}_{\beta}(-\vec{p},s).
\label{neg}\end{eqnarray}\end{mathletters}
Both sets of spinors  are eigenstates of
$\chi\equiv
\gamma_{5}\vec{\Sigma}\cdot\hat{p}=\gamma_{0}\vec{\gamma}\cdot\hat{p}$.
Since those in Eq. (\ref{pos}) have $\chi=+1$, the chirality and helicity have
the same sign  for particles and for anti-holes.
The spinors in  Eq. (\ref{neg}) have $\chi=-1$, indicating that  the chirality
and helicity have the opposite sign  for antiparticles and for holes.

\subsubsection{Additional properties}

The advanced propagator is obtained by the relation
in Eq. (\ref{relation}) of Appendix C:
\begin{eqnarray}
S^{A}(p_{0},\vec{p})&&\nonumber\\
={1\over 2}(\gamma_{0}&&-\vec{\gamma}\cdot\hat{p})
\Big({Z_{p}^{\ast}(p)\over p_{0}-{\cal E}_{p}^{\ast}}
+{Z_{h}(p)\over p_{0}+{\cal E}_{h}}-f^{\ast}(p_{0}^{\ast},p)\Big)\nonumber\\
+{1\over 2}(\gamma_{0}&&+\vec{\gamma}\cdot\hat{p})
\Big({Z_{h}^{\ast}(p)\over p_{0}-{\cal E}_{h}^{\ast}}
+{Z_{p}(p)\over p_{0}+{\cal E}_{p}}+f(-p_{0},p)
\Big).\label{A}\end{eqnarray}
It has four poles in the upper half-plane as well as branch cuts in the upper
half-plane from $f$.

As discussed in Appendix C, the spectral function may be expressed as the
difference between the retarded and advanced propagators. It is convenient to
define that part of the spectral function proportional to
$\gamma_{0}$ as
\begin{displaymath}
\rho_{0}(p_{0},p)={1\over 4}{\rm
Tr}[\,\rho(p_{0},p)\gamma_{0}].\end{displaymath}
As shown in Appendix C, $\rho_{0}$ must be positive:
\begin{eqnarray}
\rho_{0}(p_{0},p)=&&-{\rm Im}\Big[{Z_{p}(p)2{\cal E}_{p}\over p_{0}^{2}-{\cal
E}_{p}^{2}}+{Z_{h}(p)2{\cal E}_{h}\over p_{0}^{2}-{\cal
E}_{h}^{2}}\Big]\nonumber\\
 &&+{\rm Im}\Big[f(p_{0},p)+f(-p_{0},p)\Big]>0,\label{pos2}
\end{eqnarray}
 The canonical anticommutation relations of the fermion
field operators impose  require the integral of  $\rho_{0}$ to be unity
as shown in Eq. (\ref{sumrule1}). This gives
\begin{equation}
{\rm Re}\big(Z_{p}(p)\big)+{\rm Re}\big(Z_{h}(p)\big)
+\int_{-\infty}^{\infty}\!{dp_{0}\over\pi}\,{\rm Im}f(p_{0},p)=1
.\label{sumrule}\end{equation}
It is perhaps worth noting that  the explicit one-loop calculations
satisfy additional sum rules for the first and second moments of the
energy, but
these  apply only at  one-loop \cite{b3}.

The time-ordered propagator can be expressed directly in terms of the retarded
and advanced propagators as
\begin{equation}
S^{11}(p_{0},\vec{p})=S^{R}(p_{0},\vec{p})\,e^{\beta p_{0}}n(p_{0})
+S^{A}(p_{0},\vec{p})\,n(p_{0}).
\label{S11}\end{equation}
It is rather trivial to prove this for $p_{0}$ real.
Appendix D proves that it holds throughout the complex $p_{0}$
plane. Therefore in addition to the kinematic poles coming from  the
Fermi-Dirac function, the time-ordered  propagator has eight
dynamical poles: four in the lower half-plane from $S^{R}$ and four in the
upper
half-plane from $S^{A}$. Of course the dynamical poles are all reflections of
the basic particle and hole energies ${\cal E}_{p}$ and ${\cal E}_{h}$.

\subsubsection{Conclusion}

The above results depend only upon invariance under chirality, parity, charge
conjugation, and time reversal and not at all upon perturbation theory.
The  complex energies ${\cal E}_{p}(p)$ and ${\cal E}_{h}(p)$ of the
particle and
hole excitations are gauge-fixing invariant as proven generally by
Kobes, Kunsattter, and Rebhan \cite{m1}.
However the residue functions $Z_{p}$
and $Z_{h}$ are expected to change with gauge. Appendix D shows that the
renormalized electric charge for the particle and hole excitations,
as measured by the coupling to photons at zero momentum, has the correct value
independent of the functions $Z_{p}$
and $Z_{h}$.

The general shape of the two dispersion curves requires difficult
calculation. The
only simple property is that as $p\to\infty$ the effects of temperature
diminish so that ${\cal E}_{p}(p)\to p$ and $Z_{h}(p)\to 0$.
The hard thermal loop calculations enjoy several properties that have not been
proven to hold generally. Namely, in the HTL approximation the hole energy is
asymptotic to $p$ as
$p\to\infty$, the phase velocities of both excitations  are larger than
unity, and the group velocities of both excitations are smaller than unity.

At temperatures close to the critical temperature the running coupling is
 large and  one-loop calculations do not apply.
Since the doubled dispersion relation is such a characteristic feature of
the high-temperature, chirally symmetric phase of QCD, it is frustrating that
experimental signatures are so difficult to find.
One possibility
are the Van Hove singularities in dilepton production found by
Braaten, Pisarski, and Yuan \cite{d1}.
This has been recently studied by Sch\"{a}fer and Thoma \cite{l1} and
Peshier and Thoma  \cite{l1a}.
The Van Hove singularities are determined by the minimum value of the
hole dispersion relation $E_{\rm min}$ and by the maximum energy
difference between the two dispersion relations $\Delta E$.
In the dilepton rest frame  the dilepton rate has a square
root divergence at $k_{0}=2E_{\rm min}$ and at $k_{0}=\Delta E$.
Unfortunately in the HTL approximation the values $E_{\rm min}=0.38 gT$
and $\Delta E=0.19 gT$ are so small that the effect is
 swamped by the large dilepton continuum. However since the
continuum falls rapidly with energy, it is possible that the Van Hove
singularities might exceed the continuum if the true
 values of $E_{\rm min}$ and $E_{\rm max}$ are large enough. Of course,
the effect will be weakened by the quark damping rates.

\acknowledgements

It is a pleasure to thank Andre Peshier and Markus Thoma for their interest
in the
nonperturbative information that could be extracted.
This work was supported in part by the U.S. National Science Foundation
under Grant
No. PHY-9900609.

\appendix

\section{\lowercase{$p=0$ independent of $\hat{p}$}}

A crucial step in the proof\, that particles and holes have separate dispersion
relations, which is used in  Eq. (\ref{p=01}) and subsequently, is the fact
that at zero three-momentum the fermion propagator does not depend on the
momentum direction
$\hat{p}$. (This  property is specific to fermions. Gauge boson propagators
can  depend on unit vectors at zero momentum, as occurs even  in the
free Coulomb-gauge propagator.)  The argument is simple.
The self-energy at any order of perturbation theory is expressed in terms of
momentum integrations over integrands composed of  fermion propagators and
  boson propagators. The routing of the external three
momenta $\vec{p}$  through the internal propagators depends on what choice is
made for the loop momenta
$\vec{k}_{j}$ that must be integrated. Because of fermion number conservation
at each vertex, the external fermion number can  be
uniquely traced through a sequence of internal fermion propagators that form a
continuous path through any self-energy diagram.  Therefore one can
always choose the loop momenta so that  the gauge boson propagators and  the
fermion propagators in closed loops will contain various $\vec{k}_{j}$ but
never
$\vec{p}$.
 The argument of the linked fermion propagators  that connect with the external
lines will be the sum  of $\vec{p}$ with a linear combination of the
$\vec{k}_{j}$. These internal fermion propagators depend on the vector
$\vec{p}$ and not separately on $\hat{p}$ and $|\vec{p}|$. Hence if the
external
momenta
$|\vec{p}|$ is set equal to zero, there can be no dependence on the direction
$\hat{p}$ and the propagator must be as shown in Eq. (\ref{p=01}).

\section{S\lowercase{ymmetry under $p\to -p$}}

It is convenient to parameterize  $\vec{p}$ by
$p,\theta,\phi$:
\begin{eqnarray}
p_{x}=&&p\sin\theta\cos\phi\nonumber\\
p_{y}=&&p\sin\theta\sin\phi\nonumber\\
p_{z}=&&p\cos\theta.\nonumber
\end{eqnarray}
Under the transformation $p\!\to\! -p,\theta\to\pi-\theta,\phi\to
\phi+\pi$ the vector $\vec{p}$ is unchanged.
This amounts to $\hat{p}\to -\hat{p}$ and $p\to -p$.
If the propagator is considered as a function of the four variables $p_{0},
p,$ and $\hat{p}$, it must not change:
\begin{displaymath}
S^{R}(p_{0},p,\hat{p})=S^{R}(p_{0},-p,-\hat{p}).
\end{displaymath}
In terms of the definitions in Eq. (\ref{Rpropagator}) this requires
\begin{equation}
D_{-}(p_{0},p)=D_{+}(p_{0},-p),
\end{equation}
as employed in Eq. (\ref{reflectD}).

\section{R\lowercase{etarded propagator}}

A natural starting point for discussion of any finite-temperature
propagator is the
finite-temperature spectral function:
\begin{equation}
\rho_{\alpha\beta}(x)=\sum_{n}e^{-\beta E_{n}}{\langle
n|\big\{\psi_{\alpha}(x),
\overline{\psi}_{\beta}(0)\big\}|n\rangle
\over {\rm Tr}[e^{-\beta H}]},\label{rho1}\end{equation}
where the sum is over a complete set of energy eigenstates $|n\rangle$.
In the following, Dirac indices will be suppressed and $\rho$ will be treated
as a matrix. The matrix satisfies:
\begin{equation}[\rho(x)]^{\dagger}=\gamma_{0}\rho(-x)\gamma_{0}.
\label{adjoint}\end{equation}

\subsubsection*{1. Retarded vs advanced}

The basic retarded and advanced propagators are defined in coordinate space as
\begin{eqnarray}
S^{R}(x)=&&-i\theta(x_{0})\rho(x)\nonumber\\
S^{A}(x)=&& i\theta(-x_{0})\rho(x).\nonumber
\end{eqnarray}
In momentum space they have the simple dispersion relations
\begin{eqnarray}
S^{R}(p_{0},\vec{p})=&&\int_{-\infty}^{\infty}\!{d\omega\over 2\pi}
\;{\rho(\omega,\vec{p})\over p_{0}-\omega+i\eta}\label{AR}\\
S^{A}(p_{0},\vec{p})=&&\int_{-\infty}^{\infty}\!{d\omega\over 2\pi}
\;{\rho(\omega,\vec{p})\over p_{0}-\omega-i\eta}.\label{RA}\end{eqnarray}
The retarded propagator is an analytic function of complex $p_{0}$ in the
half-plane  ${\rm Im} p_{0}>0$.  Its only singularities are
in the lower half-plane.
 Similarly the advanced propagator is an analytic function of complex $p_{0}$
in the half-plane  ${\rm Im} p_{0}<0$. All
its singularities are in the upper half-plane. The two propagators are
connected by the
relation
\begin{equation}S^{A}(p_{0}^{\ast},\vec{p})
=\gamma_{0}[S^{R}(p_{0},\vec{p})]^{\dagger}\gamma_{0}\label{relation},
\end{equation}
which follows directly from the adjoint property  of the spectral
function in Eq. (\ref{adjoint}).

\subsubsection*{2. Spectral Function}

Using Eqns. (\ref{AR}) and (\ref{RA})
the spectral function in momentum space can be expressed as
\begin{equation}
\rho(\omega,p)=i\big[S^{R}(\omega,\vec{p})-S^{A}(\omega,\vec{p})\big].
\label{rho}\end{equation}
Although the spectral function is defined for real $\omega$, this relation
allows
$\omega$ to be continued into the complex plane.
From the definition of the spectral function the matrix
$[\rho(\omega,\vec{p})\gamma_{0}]_{\alpha\beta}$ is  Hermitian and has a
positive-definite trace:
\begin{equation}
\rho_{0}(\omega,\vec{p})\equiv{1\over 4}{\rm
Tr}[\rho(\omega,\vec{p})\gamma_{0}]
>0.\end{equation}
This leads to the condition in Eq. (\ref{pos2}).

From the canonical anticommutation relations at equal times,
 the spectral function must satisfy the relation
$\gamma_{0}\delta^{3}(\vec{x})=\rho(0,\vec{x}).$
The Fourier transform is
\begin{displaymath}
\gamma_{0}=\int_{-\infty}^{\infty}{d\omega\over 2\pi}
\,\rho(\omega,\vec{p}).\end{displaymath}
From Eq. (\ref{rho}) this can
be expressed as
\begin{equation}
\gamma_{0}=i\int_{-\infty}^{\infty}\!{d\omega\over
2\pi}\,\big(S^{R}(\omega,\vec{p})
-S^{A}(\omega,\vec{p})\big).\label{sum}\end{equation}
 When Eq. (\ref{R}) and (\ref{A}) are used the terms proportional to
$\vec{\gamma}\cdot\hat{p}$  automatically integrate to zero and those
proportional to
$\gamma_{0}$ give  the sum rule
\begin{equation}
\int_{-\infty}^{\infty}{d\omega\over 2\pi}\,\rho_{0}(\omega,p)=1.
\label{sumrule1}\end{equation}

\subsubsection*{3. Discrete symmetries}

In the high temperature phase, QCD  is expected to be invariant under chirality
as well as under parity, charge conjugation, and time reversal. These
symmetries constrain the spectral function and consequently the propagators.

(i) Chirality invariance requires that the spectral  function satisfy
$\rho(x)=-\gamma_{5}\rho(x)\gamma_{5}$. For the retarded propagator this
implies that
\begin{eqnarray}Q_{5}:\hskip1cm
S^{R}(p_{0},\vec{p})=-\gamma_{5}S^{R}(p_{0},\vec{p})\gamma_{5}\label{Q5}.
\end{eqnarray}

(ii) Parity requires $\rho(t,\vec{x})=\gamma_{0}\rho(t,-\vec{x})\gamma_{0}$ and
therefore
\begin{equation}
P:\hskip1cm
S^{R}(p_{o},\vec{p})=\gamma_{0}S^{R}(p_{0},-\vec{p})\gamma_{0}.\label{P}
\end{equation}

(iii) Time reversal invariance requires
$[\rho(t,\vec{x})]^{\ast}=T\rho(-t,\vec{x})T$ where
$T=i\gamma^{1}\gamma^{3}$ and relates
the retarded propagator to the advanced:
\begin{displaymath}
[S^{R}(p_{0},\vec{p})]^{\ast}=TS^{A}(p_{0}^{\ast},-\vec{p})T.\end{displaymath}
Using Eq. (\ref{relation}) this can be expressed in terms of $S^{R}$:
\begin{equation} T:\hskip0.5cm
S^{R}(p_{0},\vec{p})=T\gamma_{0}[S^{R}(p_{0},-\vec{p})]^{T}\gamma_{0}T.\label{T}
\end{equation}
This relation is automatically satisfied by Eq. (\ref{2pole}).

(iv) Charge conjugation requires $[\rho(x)]^{T}=C\rho(-x)C$ where
$C=i\gamma^{2}\gamma^{0}$. This relates the retarded propagator to the
advanced:
\begin{displaymath}
[S^{R}(p_{0},\vec{p})]^{T}=-C\,S^{A}(-p_{0},-\vec{p})\,C.
\end{displaymath}
However, using Eq. (\ref{relation}) this can be rewritten as a constraint
on the
retarded propagator directly:
\begin{equation}C:\hskip0.5cm
S^{R}(p_{0},\vec{p})=-C\gamma_{0}[S^{R}(-p_{0}^{\ast},-\vec{p})]^{\ast}
\gamma_{0}C.\label{C}\end{equation}
It is this relation that leads from Eq. (\ref{2pole}) to Eq. (\ref{R}).

(v) It is sometimes convenient to deal with the combination $\theta=TCP$, which
is always an invariance. Thus
$[\rho(x)]^{\dagger}=-\gamma_{0}\gamma_{5}\rho(x)\gamma_{5}\gamma_{0}$ always
holds and therefore
\begin{equation}
TCP:\hskip0.5cm [S^{R}(p_{0},\vec{p})]^{\dagger}
=\gamma_{5}\gamma_{0}S^{R}(-p_{0}^{\ast},-\vec{p})\gamma_{0}\gamma_{5}\label{TCP
}
\end{equation}
must always hold.

\section{T\lowercase{ime-ordered propagator}}

This appendix proves Eq. (\ref{S11}), which expresses  the
time-ordered propagator in momentum space directly in terms of the retarded
and advanced propagators. The relation holds for all $p_{0}$.
The time-ordered propagator is defined as
\begin{displaymath}
S_{\alpha\beta}^{(11)}(x)=-i\sum_{n}e^{-\beta
E_{n}}{\langle
n|T(\psi_{\alpha}(x),\overline{\psi}_{\beta}(0))|n\rangle\over{\rm
Tr}[e^{-\beta H}]}.\end{displaymath}
This can be expressed in terms of the spectral function:
\begin{displaymath}
S^{11}(p_{0},\vec{p})=\int_{-\infty}^{\infty}\!{d\omega\over
2\pi}\Big({\rho(\omega,\vec{p})
\,[1-n(\omega)]\over p_{0}+i\eta-\omega}+{\rho(\omega,\vec{p})
\,n(\omega)\over p_{0}-i\eta-\omega}\Big),
\end{displaymath}
where $n(\omega)$ is the Fermi-Dirac distribution function:
\begin{displaymath}
n(\omega)={1\over\exp(\beta\omega)+1}.\end{displaymath}
It is rather easy to evaluate $S^{11}$ for $p_{0}$ real and $\eta$
infinitesimal
and obtain Eq. (\ref{S11}) for real $p_{0}$. However, since  the focus of the
present paper is  the pole structure of the propagators in the complex $p_{0}$
plane, that easy argument is  not adequate. In order for Eq. (\ref{S11})
to hold in the complex $p_{0}$ plane it will be important that
 $n(\omega)$  have not branch points on the
real axis.

The dispersion relation for  $S^{11}$ shows it to be  the sum of two
functions. The
first
 is analytic in ${\rm Im} p_{0}>0$ and the second is analytic in
${\rm Im} p_{0}<0$. Write $S^{11}$ as
\begin{displaymath}
S^{11}(p_{0},\vec{p})=S^{R}(p_{0},\vec{p})+F(p_{0},\vec{p})+
G(p_{0},\vec{p}),
\end{displaymath}
where
\begin{eqnarray}
{\rm Im}\,p_{0}>0:\hskip0.5cm
F(p_{0},\vec{p})=&&-\int_{-\infty}^{\infty}{d\omega\over 2\pi}
{\rho(\omega,\vec{p})\,n(\omega)\over p_{0}-\omega}\nonumber\\
 {\rm Im}\,p_{0}<0:\hskip0.5cm
G(p_{0},\vec{p})=&&\int_{-\infty}^{\infty}{d\omega\over 2\pi}
{\rho(\omega,\vec{p})\,n(\omega)\over p_{0}-\omega}.\nonumber
\end{eqnarray}
Using Eqns. (\ref{rho}) for the spectral function, $F$ can be computed by
closing
the $\omega$ contour in the upper half-plane for the term containing
$S^{R}$ and
closing it in the lower half-plane for the term containing $S^{A}$. Since
$n(\omega)$ has poles at $\omega=\pm\Omega_{\ell}$ where
$\Omega_{\ell}=i\ell\pi T$ the result for $F$ is
\begin{displaymath}
-S^{R}(p_{0},\vec{p})n(p_{0})-T\sum_{\ell=1}^{\infty}
\Big({S^{R}(\Omega_{\ell},\vec{p})
\over p_{0}-\Omega_{\ell}}+
{S^{A}(-\Omega_{\ell},\vec{p})
\over p_{0}+\Omega_{\ell}}\Big).
\end{displaymath} This is analytic for ${\rm Im}\,p_{0}>0$ but has poles and
cuts for ${\rm Im}\,p_{0}<0$. In the same fashion $G$ can be computed by
closing the $\omega$ contour in the upper half-plane for the term containing
$S^{R}$ and closing  in the lower half-plane for the term containing $S^{A}$
with the  result
\begin{displaymath}
T\sum_{\ell=1}^{\infty}
\Big({S^{R}(\Omega_{\ell},\vec{p})
\over p_{0}-\Omega_{\ell}}+
{S^{A}(-\Omega_{\ell},\vec{p})
\over p_{0}+\Omega_{\ell}}\Big)+S^{A}(p_{0},\vec{p})n(p_{0}).
\end{displaymath}
This is analytic for ${\rm Im}\,p_{0}<0$ but has poles and cuts in ${\rm
Im}\,p_{0}>0$.  When the results for $F$ and $G$ are added the sums over $\ell$
coming from the poles of the  Fermi-Dirac function $n(\omega)$ cancel and give
\begin{displaymath}
F(p_{0},\vec{p})+G(p_{0},\vec{p})=\big(-S^{R}(p_{0},\vec{p})
+S^{A}(p_{0},\vec{p})
\big)\,n(p_{0}).\end{displaymath}
The final result for the time-ordered propagator is
\begin{displaymath}
S^{11}(p_{0},\vec{p})=S^{R}(p_{0},\vec{p})\,[1-n(p_{0})]
+S^{A}(p_{0},\vec{p})\,n(p_{0}).
\end{displaymath}
Since $1-n(p_{0})=e^{\beta p_{0}}n(p_{0})$ the result can be expressed as
 Eq. (\ref{S11}).

\section{E\lowercase{lectric charge}}

This appendix will show that the  four poles in the retarded propagator
(\ref{R}) all have the correct value of the renormalized electric charge.
The operator  $\psi$ for the $u$ quark destroys charge $e_{\psi}=2e/3$;
the operator  $\psi$ for the $d$ quark destroys charge $e_{\psi}=-e/3$.
The amplitude for a photon  with zero four-momentum to couple to a fermion is
\begin{displaymath}
S_{R}(p_{0},\vec{p})\Gamma_{\mu}(p_{0},\vec{p})S_{R}(p_{0},\vec{p}).
\end{displaymath}
At finite temperature the  Ward identity
\begin{displaymath}
\Gamma_{\mu}={\partial\over\partial
p^{\mu}}\big[S^{R}(p_{0},\vec{p})\big]^{-1}.
\end{displaymath}
determines the photon vertex function \cite{W1,W2,W3}.
The inverse of the propagator in Eq. (\ref{Rpropagator}) can be written
\begin{displaymath}
[S(p_{0},\vec{p})]^{-1}=\gamma_{0}A(p_{0},p)-\vec{\gamma}\cdot\hat{p}B(p_{0},p),
\end{displaymath}
where $D_{\pm}=A\mp B$.
The $\Gamma_{0}$ vertex  is thus
\begin{equation}
\Gamma_{0}(p_{0},\vec{p})=\gamma_{0}A'(p_{0},p)
-\vec{\gamma}\cdot\hat{p}B'(p_{0},p),\label{Gamma0}\end{equation}
where prime denotes the $p_{0}$ derivative.

\subsubsection*{1. Pole at ${\cal E}_{p}$ or $-{\cal E}_{h}^{\ast}$}

When $p_{0}\to {\cal E}_{p}$ the retarded propagator, Eq. (\ref{R}), is
dominated
by the pole
\begin{displaymath}
S^{R}(p_{0},\vec{p})\to Z_{p}(p){\sum_{\rm spin}u(\vec{p})\overline{u}(\vec{p})
\over p_{0}-{\cal E}_{p}}.
\end{displaymath}
If the quark field $\psi$ carries bare electric charge $e_{0\psi}$,
then the   renormalized electric charge which couples the photon to the
particle excitation is
\begin{displaymath}
Q=e_{0\psi}\sqrt{Z_{3}}Z_{p}(p)\cdot\overline{u}(\vec{p})\Gamma_{0}(p_{0},\vec{p
})
u(\vec{p}),
\end{displaymath}
where $\sqrt{Z_{3}}$ comes from the photon propagator and $\sqrt{Z_{p}}$ comes
from each of the two fermion fields at the vertex.
Substituting Eq. (\ref{Gamma0}) gives
\begin{displaymath}
Q=e_{0\psi}\sqrt{Z_{3}}Z_{p}(p)[A'(p_{0},p)-
B'(p_{0},p)\big]_{p_{0}={\cal E}_{p}}.
\end{displaymath}
As $p_{0}\to {\cal E}_{p}$ the denominator $D_{+}=A-B$ vanishes linearly:
$A-B\to
(p_{0}-{\cal E}_{p})/Z_{p}(p)$, so that
\begin{equation}
Q=e_{0\psi}\sqrt{Z_{3}}=e_{\psi}.\label{pcharge}\end{equation}
It is easy to check that the pole in Eq. (\ref{R}) at $p_{0}=-{\cal
E}_{h}^{\ast}$ also couples with the same  charge (\ref{pcharge}).

\subsubsection*{2. Pole at ${\cal E}_{h}$ or $-{\cal E}_{p}^{\ast}$}

At the hole excitation,  $p_{0}\to {\cal E}_{h}$, the retarded propagator,
Eq. (\ref{R}), behaves as
\begin{displaymath}
S^{R}(p_{0},\vec{p})\to Z_{h}(p){\sum_{\rm
spin}v(-\vec{p})\overline{v}(\vec{-p})
\over p_{0}-{\cal E}_{h}}.
\end{displaymath}
The renormalized electric charge of the hole excitation is
\begin{displaymath}
Q=e_{0\psi}\sqrt{Z_{3}}Z_{h}(p)\cdot\overline{v}(-\vec{p})\Gamma_{0}(p_{0},\vec{
p})
v(-\vec{p}).
\end{displaymath}
Substituting from Eq. (\ref{Gamma0}) gives
\begin{displaymath}
Q=e_{0\psi}\sqrt{Z_{3}}Z_{h}(p)\big[A'(p_{0},p)+B'(p_{0},p)
\big]_{p_{0}={\cal E}_{h}}.
\end{displaymath}
At the hole excitation, $D_{-}=A+B\to (p_{0}-{\cal E}_{h})/Z_{h}(p)$
as $p_{0}\to {\cal E}_{h}$
 so that
\begin{equation}
Q=e_{0\psi}\sqrt{Z_{3}}=e_{\psi}.\end{equation}
The same argument applies to the pole in Eq. (\ref{R}) at
$p_{0}=-{\cal E}_{p}^{\ast}$.

\end{document}